\documentclass[aps,prd,twocolumn,preprintnumbers,superscriptaddress,nobibnotes,floatfix,longbibliography]{revtex4-1}

\pdfoutput=1
\usepackage{amsmath,amsfonts,amssymb,mathrsfs,graphicx,longtable,bm,bbm}
\usepackage{array}
\usepackage{xcolor}
\usepackage[utf8]{inputenc}
\usepackage[bookmarks, pagebackref=false]{hyperref}

\usepackage{hyperref}
\definecolor{rossoCP3}{cmyk}{0,.88,.77,.40}
\definecolor{blaa}{rgb}{0.2,0.2,0.6}
\hypersetup{colorlinks, 
    bookmarksopen, 
    bookmarksnumbered,
    citecolor=blaa,
    linkcolor=rossoCP3,
    urlcolor=rossoCP3, 
}

\newcommand{\MDM}{M_{\text{DM}}}

\newcommand{\be}{\begin{equation}}
\newcommand{\ee}{\end{equation}}

\graphicspath{{./figs/}}

\makeatletter

\def\hhref#1{\href{http://arxiv.org/abs/#1}{arXiv:#1}}
\usepackage{xstring} 
\newcommand{\hhrefq}[1]{\IfSubStr{#1}{:}{\href{http://inspirehep.net/search?ln=en&ln=en&p=#1&of=hb&action_search=Search&sf=&so=d&rm=&rg=25&sc=0}{InSpires:#1}}{\hhref{#1}}}

\def\art{\@ifnextchar[{\eart}{\oart}}
\def\eart[#1]#2#3#4#5#6{{\rm #2}, {\em #3 \bf #4} {\rm (#6) #5} ({\em #1})}
\def\article{\@ifnextchar[{\earticle}{\oarticle}}
\def\oarticle#1#2#3#4#5#6{{\rm #1}, {``#6''}, {\rm #2 #3 (#5) #4}}
\def\earticle[#1]#2#3#4#5#6#7{{\rm #2}, {``#7''}, {\rm #3 #4 (#6) #5}  [\hhrefq{#1}]}
\def\hepart[#1]#2{{\rm #2, \sl#1}}
\def\heparticle[#1]#2#3{#2, { ``#3''} [\hhrefq{#1}]}

\begin{document}


\title{\Large\color{rossoCP3} Dark Matter meets Quantum Gravity}
\author{Manuel {\sc Reichert}}
\thanks{{\scriptsize Email}: \href{mailto:reichert@cp3.sdu.dk}{reichert@cp3.sdu.dk}; {\scriptsize ORCID}: \href{https://orcid.org/0000-0003-0736-5726}{ 0000-0003-0736-5726}}
\affiliation{CP$^3$-Origins,  University of Southern Denmark, Campusvej 55, 5230 Odense M, Denmark}
\author{Juri {\sc Smirnov}}
\thanks{{\scriptsize Email}: \href{mailto:smirnov.9@osu.edu}{smirnov.9@osu.edu}; {\scriptsize ORCID}: \href{https://orcid.org/0000-0002-3082-0929}{ 0000-0002-3082-0929}}
\affiliation{Center for Cosmology and AstroParticle Physics (CCAPP), The Ohio State University, Columbus, OH 43210, USA}
\affiliation{Department of Physics, The Ohio State University, Columbus, OH 43210, USA}

\begin{abstract}
We search for an extension of the Standard Model that contains a viable dark matter candidate and that can be embedded into a fundamental, asymptotically safe, quantum field theory with quantum gravity.
Demanding asymptotic safety leads to boundary conditions for the non-gravitational couplings at the Planck scale.
For a given dark matter model these translate into constraints on the mass of the dark matter candidate. 
We derive constraints on the dark matter mass and couplings in two minimal dark matter models:
i) scalar dark matter coupled via the Higgs portal in the $B$-$L$ model;
ii) fermionic dark matter in a $U(1)_X$ extension of the Standard Model, coupled via the new gauge boson.
For scalar dark matter, we find 56\,GeV $ <  \MDM < 63$\,GeV, and for fermionic dark matter, $\MDM \leq 50$\,TeV.
Within our framework, we identify three benchmark scenarios with distinct phenomenological consequences. 
\\
[.3cm]
{\footnotesize  \it Preprint: CP$^3$-Origins-2019-41 DNRF90
}

\end{abstract}

\maketitle

\section{Introduction}
Our current understanding of nature demands the existence of additional matter degrees of freedom. Separately, quantum effects from the gravity sector must affect our quantum field theory (QFT) framework at energies close to the Planck scale.  In this paper, we simultaneously address these two questions and explore what the high energy effects of gravity imply for the new matter degrees of freedom at the low energy scale.  Our working hypothesis is that the underlying QFT, which contains a dark matter candidate, should become asymptotically safe with the inclusion of quantum gravity. This sets constraints on the model parameter space leading to predictions for dark matter phenomenology.

Since our best current description of microscopic processes in nature is QFT, we will extend the current theory that describes the physics of the visible sector, the Standard Model (SM), by additional quantum fields. The new fields have to be stable and account for the dark matter (DM) component of our Universe. In our extensions, we are guided by minimality, which naturally leads us to consider the simplest known production mechanism for such dark sector particles, the thermal freeze-out~\cite{Zeldovich:1965gev, Lee:1977ua,Steigman:1984ac,Kolb:1990vq, hep-ph/9506380,Bertone:2004pz,1204.3622,Arcadi:2017kky,Roszkowski:2017nbc}. 

The question we address in this paper is which minimal models with a DM candidate have an ultraviolet (UV) safe embedding into a theory of quantum gravity and what that implies for their available parameter space. While the observations of galaxies and clusters can have substantial uncertainties, when it comes to predicting the exact value of the missing DM component~\cite{1809.05318}, the observations of the cosmic microwave background lead to a very solid measurement of its abundance, which we use as our benchmark $\Omega_\text{DM} h^2 \approx 0.12$~\cite{Aghanim:2018eyx}. Previous work has found DM mass bounds from the general consideration of unitarity~\cite{Griest:1989wd,1904.11503}, our bounds lead to more stringent mass constraints. 

We choose a minimal approach to quantum gravity, assuming the framework of QFT and no additional degrees of freedom besides the spin-2 field in the gravitational sector. 
As pointed out in Ref.~\cite{Weinberg:1980gg}, the UV behavior of a QFT describing quantum gravity might be governed by a non-trivial fixed point.
This UV fixed point would make the theory UV finite and thus non-perturbatively renormalizable.
Starting with the seminal work by Reuter \cite{Reuter:1996cp},
a lot of evidence was collected in favor of this scenario 
\cite{hep-th/0110054,0805.2909,1211.0955,1404.4537,1410.4815,1504.07656,1506.07016,1507.08859,1601.01800,1607.08460,1609.04813,1612.07315,1711.09259,1810.08550}.
The interplay of quantum gravity and matter was extensively investigated as well \cite{1311.2898,Meibohm:2015twa,1702.06539,Eichhorn:2018akn,1802.00498,1803.02355,Eichhorn:2018ydy,1812.08782,1907.02903}.

In a fundamental theory of nature not only must the gravity couplings become asymptotically safe,
but the matter coupling must also be either asymptotically safe or free.
Due to this requirement, asymptotically safe quantum gravity can, in some cases, predict the values of couplings in the SM.
These predictions appear as boundary conditions at the Planck scale.
If these boundary conditions are not fulfilled then couplings typically run into Landau poles.

The first prediction of asymptotically safe quantum gravity was the Higgs boson mass \cite{Shaposhnikov:2009pv}:
asymptotic safety predicts that the quartic scalar coupling is roughly vanishing at the Plank scale.
This yields a Higgs boson mass in the range from roughly 126 to 136\, GeV, depending, for instance, on the value of the top mass.
Also, a retrodiction of the top mass \cite{Eichhorn:2017ylw} and the difference between the top and the bottom mass \cite{Eichhorn:2018whv} were attempted.
See also \cite{1709.07252,1711.02949,1810.07615,1909.07318} for further works in this context.

This paper is structured as follows:
In \autoref{sec:executive-summary} we give a short executive summary of our main ideas and explain why we consider certain dark models.  
In \autoref{sec:QG}, we detail in detail the quantum gravity contribution to the beta functions of the matter couplings and how they lead to boundary conditions at the Planck scale.
In \autoref{sec:DM}, we present the DM model that we consider in two different mass hierarchies, in one mass hierarchy the scalar and in the other, the fermion is the DM candidate. 
In both cases, we show how the boundary condition from gravity is applied and the consequences on the computed relic density.
In \autoref{sec:discussion}, we critically discuss our findings, in particular, the uncertainty in the computation of the quantum gravity contributions.

\section{Executive summary for\\ the busy reader}
\label{sec:executive-summary}
We work under the hypothesis that QFT is a fundamental description of nature at all scales.  Thus, we have to take into account the effects of quantum gravity when approaching the Planck scale. Extensive research in this direction has been conducted and we discuss the technical aspects in the following section. However, the main point is, that quantum gravity provides Planck scale boundary conditions for the renormalization group (RG) flow equations. We will demonstrate that those conditions constrain the allowed masses of DM candidates for the simplest models of DM. 

Our approach is based on several key assumptions. The first assumptions are that the RG flow of all matter fields remains stable up to the Planck scale. Thus no Landau poles or vacuum instabilities occur below Planckian energies. As we will see, this assumption alone provides constraints on the allowed DM models and is independent of the assumed theory of quantum gravity.  Furthermore, we assume the following about the theory of quantum gravity
\begin{itemize}
\item Spacetime is $3+1$ dimensional.
\item The finiteness of the gravitational and matter couplings is guaranteed by an asymptotically safe fixed point. 
\item The transition from the classical gravity regime to the asymptotically safe regime happens close to the Planck scale.
\end{itemize}
The description of quantum gravity is minimal in the sense that it does not introduce new concepts nor new fields in the gravitational sector. The entire system including gravity is described in the framework of quantum field theory. 

These assumptions lead us to boundary conditions for the matter couplings at the Planck scale.
Our results also hold if similar boundary conditions are obtained from different assumptions.
For example, scale invariance above the Planck scale in the scalar sector and the demand that all gauge couplings remain perturbative until the Planck scale, lead to similar boundary conditions.  

For a theory to contain a DM candidate, a new field has to be present which 
\begin{itemize}
\item is stable or long-lived on cosmic time scales.
\item has a portal interaction with the SM fields in order to be produced in the early Universe.
\end{itemize}
Among the simplest portals to the dark sector is the Higgs field. The renormalizable interaction $ \lambda_p \,H^\dagger H\, S\,S^*$ is unavoidable once a new scalar field is present in the theory and it can communicate between the SM and the dark sector. The interaction strength is controlled by the portal coupling $\lambda_p$. This portal coupling is forced to be roughly zero at the Planck scale by the quantum gravity contributions \cite{Eichhorn:2017als,Pawlowski:2018ixd}.

In Ref.~\cite{Eichhorn:2017als} it has been argued that this portal setup is not viable if there are only scalars in the dark sector.
The reason is that the interaction parameter is multiplicatively renormalized and thus not generated once set to zero.
We explore a dark sector where also other interactions are present that can generate the portal coupling. What can those interactions be?

One possibility is a Yukawa induced portal. Here a new fermion with interactions to the DM scalar $y_s \bar{\psi} \psi S$ and the Higgs boson $y_h \bar{\psi} \psi H$ can generate the portal coupling at one loop. However, this interaction breaks the $Z_2$ symmetry, which is essential for the stability of the scalar field $S$. Consequently, this scenario does not lead to a long-lived field $S$.

The other possibility is the gauge induced portal. We argued that the portal coupling has to be induced by an interaction that preserves the stabilizing $Z_2$ symmetry. This can be the case if a new gauge force is present in nature. The new gauge boson has to couple to the DM scalar field and at the same time couple to the Higgs scalar. This can be realized in two ways, either through a quantum number assignment to the new gauge boson, which contains hypercharge, or kinetic mixing to the $U(1)_Y$ gauge boson. 

The new gauge symmetry can remain unbroken if the gauge boson has a Stueckelberg type mass~\cite{1506.05107}, or be spontaneously broken at a higher scale. In either case, by an appropriate choice of quantum numbers, a stable field naturally arises in the theory. This field can either be: 
\begin{itemize}
\item a scalar field $S$, with an induced Higgs portal coupling. 
We perform the RG analysis in the case where the new $U(1)$ symmetry is the $B$-$L$ symmetry.
This gauge symmetry is the simplest way to make the additional heavy fermions decay in order not to be overproduced. Those fermions are a necessary ingredient to guarantee the vacuum stability of the scalar field $S$. 
We find an upper bound on the scalar portal coupling $\lambda_p$ of the order of $10^{-1}$ at the DM scale. The relic density constraint, in this case, can only be satisfied if the DM mass is close to the Higgs resonance, which implies that $\MDM \approx m_h/2$. 
\item a new fermion field, which couples to the SM through the gauge boson portal ($Z'$) of the new gauge symmetry. Since the value of the gauge coupling at the DM scale is bounded from above due to the high-scale boundary condition, we can derive an upper bound on the DM mass. We find the upper bound on the DM mass to be $\MDM < 50$\,TeV in the $U(1)_X$ gauge extension of the SM. Note that the relic density requirement in this maximal mass scenario is only satisfied if the annihilation cross section is resonantly enhanced. 
\end{itemize}

If we additionally require that all scalar quartic couplings remain positive and no vacuum instabilities arise, we are also forced to introduce heavy fermions in the scalar DM model. We thus argue that both DM scenarios are realizations of the same model with different hierarchies. In the first case, the lightest dark sector particle is a light scalar, without a vacuum expectation value (vev), and in the second case a light fermion, while the heavier scalar can also get a vev. 

In summary, we analyze the RG flow in asymptotically safe quantum field theories with a symmetry structure that permits long-lived relics. We find surprisingly low upper bounds on the masses of those fields. 
The predicted masses are well within the reach of current or near-future indirect and direct DM searches, even though in the extremely resonant scenarios a detection is more challenging~\cite{1508.04418,1510.07562}.

\section{Quantum gravity contributions to the beta functions}
\label{sec:QG}
Graviton fluctuations alter the running of all matter couplings.
Below the Planck mass, they are strongly suppressed and thus negligible. 
Beyond the Planck scale, the contributions become strong and lead to a significantly different running of the couplings compared to the SM.
Depending on the sign of the contributions they could either prevent or trigger Landau poles, and prevent or assist asymptotic freedom.
For example, the $U(1)$-gauge coupling runs into a Landau pole without graviton fluctuations beyond the Planck scale.
Studies suggest that graviton fluctuations are strong enough to prevent that Landau pole \cite{1101.6007,1702.07724,1709.07252}. 

Gravity couples universally to all matter fields.
This means that quantum gravity contributes to the running of all gauge couplings with the same strength, independent of the gauge group.
The same holds for all Yukawa couplings and quartic scalar couplings.  
This allows us discussing general features that such couplings have in the regime beyond the Planck scale.
In the following subsections, we will detail this for each coupling separately.  

The suppression of the graviton fluctuations below the Planck scale $M_\text{Pl}$ is described by threshold functions.
They are roughly given by $\mu^2 /(M_\text{Pl}^2+ \tilde G^* \mu^2)$, where $\mu$ is the RG scale and $\tilde G^*$ the dimensionless fixed-point value of the Newton coupling.
In this work, we model the suppression with a Heaviside function for simplicity,
i.e., we model the threshold function as $\Theta(\mu^2-M_\text{Pl}^2)$.
The error introduced by this approximation is negligible compared to the general uncertainty of the graviton contributions.
This approximation allows us to use the standard perturbative beta functions without gravity below the Planck scale, 
while the boundary conditions for the matter couplings at the Planck scale are determined with gravity.

The quantum gravity contributions are obtained with a non-perturbative computation via the functional renormalization group \cite{Wetterich:1992yh}, see also \cite{Ellwanger:1993mw,Morris:1993qb}.
These contributions depend on all gravitational couplings, including the Newton coupling $G$, the cosmological constant $\Lambda$ as well as higher derivative couplings.
Examples for the higher derivative couplings are the couplings associated with $R^2$ and $R_{\mu\nu}^2$.
In the present work, we treat these contributions as just a number $f_i$.
In the regime beyond the Planck scale, these indeed become constant.
We do not need the dependence on the gravity couplings,
since we are only interested in the boundary conditions at the Planck scale.
We extract the values of these numbers from previous computations as detailed in the next sections.
There is theoretical uncertainty in the numerical values of the $f_i$ and we vary them to estimate the uncertainty of our predictions.

Non-perturbative quantum gravity computations are scheme dependent and often performed in an Einstein-Hilbert truncation. 
Nevertheless, the $f_i$ contain physical information once a particular scheme is fixed, and we can use them to determine the physical boundary conditions for a given truncation.
See \autoref{sec:discussion}, for further discussion of the uncertainties.

The kind of boundary condition at the Planck scale depends on whether a given coupling is 
\begin{itemize}
\item UV attractive (relevant) at a fixed point,
\item UV repulsive (irrelevant) at a fixed point.
\end{itemize}
For a UV attractive direction, all trajectories in the vicinity of the fixed point lead in the UV direction towards it.
For a UV repulsive direction, only one trajectory leads to the fixed point.
Consequently, an attractive direction has a range of coupling values that lead to the fixed point, while a repulsive direction has only one.
Notably, UV repulsive directions have a higher predictive power.  
The linearized flow equations around the fixed point determine whether a direction is attractive or repulsive.
More precisely, positive eigenvalues of the stability matrix ($B_{ij}=-\partial_{g_i} \beta_{g_j}$) belong to UV attractive directions,
while negative eigenvalues belong to UV repulsive directions.

\subsection{Quartic scalar coupling}
We discus the graviton contributions to a quartic scalar self coupling with a Lagrangian of the type
\begin{align}
\mathcal L \sim |D_\mu \phi|^2 + m_\phi^2 |\phi|^2 + \lambda |\phi|^4 \,.
\end{align}
The following conclusions hold independent of whether $\phi$ is a real or complex scalar field, whether it has gauge interactions or not.
Gravity contributions to this system were computed, e.g., in \cite{Percacci:2003jz,Rodigast:2009zj,Zanusso:2009bs,Narain:2009fy,Eichhorn:2017als,Pawlowski:2018ixd}.

We split the beta function in a part that stems from matter fluctuations $\beta_{\lambda,\text{matter}}$ and in a part that stems from the graviton fluctuations $f_\lambda$:
\begin{align} \label{eq:beta-lambda}
 \beta_{\lambda} = \beta_{\lambda,\text{matter}} + f_\lambda  \lambda \,.
\end{align}
In \cite{Pawlowski:2018ixd}, the contribution $f_\lambda$ was computed in an Einstein-Hilbert like truncation:
\begin{align}
 f_\lambda ={}& \frac{1}{8 \pi}  \tilde G \left( \frac{20}{(1 - 2 \tilde \Lambda)^2} + \frac{1}{(1- \tilde \Lambda/2)^2} \right) \notag \\
 &+ \frac{1}{8 \pi}  \frac{\tilde G  \tilde m_\phi^4}{\lambda} \left( \frac{80}{(1 - 2 \tilde \Lambda)^3} + \frac{1}{(1 - \tilde \Lambda/2)^3} \right)  \,,
\end{align}
where $\tilde G=G \mu^2$, $\tilde \Lambda =\Lambda/\mu^2$,  and $\tilde m_\phi = m_\phi/\mu$ are the dimensionless versions of the Newton coupling, cosmological constant and scalar mass, respectively, and $\mu$ is the RG scale.
Importantly, the gravitational contribution allows for a Gau\ss ian fixed point $\lambda^* =\tilde m_\phi^*=0$,
which is also a fixed point of $ \beta_{\lambda,\text{matter}}$, assuming that the gauge and Yukawa couplings are vanishing.
Indeed the Gau\ss ian fixed point was found to be the only fixed point of the system \cite{Eichhorn:2017als,Pawlowski:2018ixd}.
The fixed point becomes almost Gau\ss ian, if the gauge and Yukawa couplings are not vanishing, 
typically with a small negative value for the quartic coupling, $\lambda^*\approx 0$.

The predictive quality of the quartic scalar coupling \cite{Shaposhnikov:2009pv}
stems from the fact that it is UV repulsive at this (almost) Gau\ss ian fixed point \cite{Eichhorn:2017als,Pawlowski:2018ixd}.
This entails that only one trajectory leads to the fixed point and the coupling value is fully determined at the Planck scale.
This leads to the prediction 
\begin{align}
 \label{eq:prediction-lambda}
 \lambda (M_\text{Pl}) \approx 0 \,.
\end{align}
The same boundary condition is, for example, also obtained in the 'flatland scenario' \cite{1310.4304,1401.5944}.

\subsection{Gauge coupling}
\label{sec:gauge}
We now discuss the graviton contribution to the running of the gauge coupling,
which was computed, e.g., in \cite{Daum:2009dn,1101.6007,Folkerts:2011jz,1702.07724,Christiansen:2017cxa,1709.07252}.
We again split the beta function in the standard matter part $\beta_{g,\text{matter}}$ and into a gravity part $f_g g$:
\begin{align}
 \label{eq:beta-g}
 \beta_g = \beta_{g,\text{matter}} - f_g g \,.
\end{align}
The contribution $f_g$ does not depend on the type of gauge symmetry.
In Ref.~\cite{Christiansen:2017cxa} it was computed with the result 
\begin{align}
\label{eq:fg}
f_g  = \frac{\tilde G}{16\pi}\left( \frac{8}{1-2 \tilde \Lambda} - \frac{4}{(1-2 \tilde \Lambda)^2} \right) \,.
\end{align}
Again, $\tilde G=G \mu^2$ and $\tilde \Lambda =\Lambda/\mu^2$ are the dimensionless versions of the Newton coupling and cosmological constant.
Importantly, $f_g$ is positive for all relevant values of the gravity couplings, see \cite{Christiansen:2017cxa},
which makes the contribution to the beta function negative.
Typical values of $f_g$ are of the order $\mathcal{O}(10^{-2})$ \cite{Eichhorn:2018whv}
and here we use $f_g \leq 0.04$.

Gravity supports asymptotic freedom for non-Abelian gauge theories and the gauge couplings flow into the Gau\ss ian fixed point $g^*=0$~\cite{Folkerts:2011jz,Christiansen:2017cxa}.
These directions of the Gau\ss ian fixed point are relevant and thus the gauge couplings approach it slowly beyond the Planck scale. No prediction can be made for their values at the Planck scale.

For Abelian gauge theories and asymptotically non-free non-Abelian gauge theories, this negative contribution can prevent the Landau pole of the gauge coupling \cite{Daum:2009dn,1709.07252}.
To be more precise, if the graviton contributions are strong enough compared to the strength of the gauge coupling,
then the Landau pole is avoided and the gauge coupling becomes either asymptotically free or safe.
For a given gravity contribution $f_g$ this results in an upper bound for the gauge coupling at the Planck scale.
For example if we look at the beta function of the gauge coupling at one loop
\begin{align}
 \beta_g = \beta_{g,\text{1-loop}}\, g^3 - f_g g \,,
\end{align}
then the upper bound for the gauge coupling at the Planck scale is given by
\begin{align}
 \label{eq:prediction-g}
 g(M_\text{Pl}) \leq \sqrt{\frac{f_g}{\beta_{g,\text{1-loop}}}}\,.
\end{align}

\subsection{Yukawa coupling}
\label{sec:yukawa}
The story for Yukawa couplings is similar to the Abelian gauge coupling.
The gravitational contribution needs to be negative to be phenomenologically viable.
This has been extensively discussed in Refs.~\cite{Rodigast:2009zj,Zanusso:2009bs,Oda:2015sma,Eichhorn:2016esv,1705.02342}.
The negative contribution leads to a UV attractive Gau\ss ian fixed point and a UV repulsive interacting fixed point.
In combination, this yields an upper bound for the Yukawa couplings at the Planck scale.

We split the beta function into its contributions and look only at the one-loop contribution in the matter sector
\begin{align}
 \label{eq:beta-y}
 \beta_y = \beta_{y,\text{1-loop-yukawa}} y^3 - \beta_{y,\text{1-loop-gauge}} y  - f_y y \,,
\end{align}
where $\beta_{y,\text{1-loop-gauge}}$ is positive and depends on the gauge couplings.
Then the upper bound at the Planck scale is given by
\begin{align}
 \label{eq:prediction-y}
 y(M_\text{Pl}) \leq \sqrt{\frac{f_y + \beta_{y,\text{1-loop-gauge}}}{\beta_{y,\text{1-loop-yukawa}}}}\,.
\end{align}
Note again, that $\beta_{y,\text{1-loop-gauge}}$ depends on the gauge couplings, which might go to zero quickly.
Hence, the true bound on the Yukawa coupling might be even tighter.

\subsection{Summary of predictivity}
\label{sec:summary-predictivity}
In summary, the asymptotic safety scenario for quantum gravity leads to the boundary conditions at the Planck scale displayed in \eqref{eq:prediction-lambda}, \eqref{eq:prediction-g} and \eqref{eq:prediction-y}.
There is no boundary condition for non-Abelian gauge couplings, assuming that they are asymptotically free by themselves.
Below the Planck scale, the contributions from graviton fluctuations are strongly suppressed
and we compute the running of the couplings with standard perturbative beta functions.

The existence of boundary conditions at the Planck scale raises the question of the compatibility of the SM couplings at the Planck scale with their known values at the electroweak scale.
The hypercharge and the Yukawa couplings would not be compatible if the values for $f_g$ and $f_y$ in \eqref{eq:prediction-g} and \eqref{eq:prediction-y} were too small.
This results in lower limits $f_g \geq 9.8\cdot10^{-3}$and $f_y\geq 10^{-4}$, assuming only SM matter content \cite{Eichhorn:2018whv}.
These values are in agreement with non-perturbative computations, see, e.g., \eqref{eq:fg}.

The critical coupling is the quartic Higgs coupling $\lambda_h$.
Fixing the Higgs mass to its observed value and using the SM RG running, one obtains a prediction for the quartic Higgs coupling which is slightly negative at the Planck scale $\lambda_h(M_\text{Pl})= -0.0143$ for a top pole mass of $m_t=173$\,GeV \cite{Buttazzo:2013uya}.
On the other hand, if we fix $\lambda_h(M_\text{Pl})\approx 0$ and use the SM RG running down to the electroweak scale, the Higgs mass is $m_h\approx130$\,GeV using two-loop RG equations and $m_h\approx136$\,GeV with one-loop RG equations, compared to the experimental value of $m_h= 125$\,GeV.
It has to be emphasized, that this computation still has some uncertainty due to the uncertainty of the top mass,
and also extensions of the SM do influence the value of the Higgs mass, see, e.g., \cite{Kwapisz:2019wrl}.
Indeed, in the latest measurements the top pole mass was determined with $m_t=171\pm 1$\,GeV~\cite{1904.05237,1905.02302},
which hints towards a smaller tension between the UV and the IR value of the quartic Higgs coupling.

In order to investigate the constraints for SM extensions with DM, we demand that the Higgs mass prediction should not be significantly worse than in the SM alone. 
That means that using one-loop RG equations, the resulting Higgs mass should lie in the interval $m_h=(125\pm 10)$\,GeV.

\section{Dark Matter models}
\label{sec:DM}
For a successful DM model, we need to generate portal interactions of the dark sector with the SM and preserve DM stability.
The basic realization is an interaction that respects a $Z_2$ symmetry. The simplest such interaction is provided by an Abelian gauge field. 
Furthermore, the SM extension has to show stable RG trajectories and not feature low lying Landau poles or vacuum instabilities. Thus, we are naturally led to an extension that mimics the SM particle content in the sense that it is a gauge-Yukawa theory.  

The new symmetry, let us call it $U(1)_X$, radiatively generates the scalar portal below the Planck scale.
It is important to include the kinetic mixing between the hypercharge $U(1)_Y$ group and the dark $U(1)_X$ group.
The kinetic mixing guarantees that the scalar portal is generated even if the Higgs is not charged under the new symmetry.

The Lagrangian of the dark sector reads 
\begin{align}
\label{eqn:dmmodel}
&  \mathcal L_D \sim  \mathcal{L}_{\rm scalar} +   \mathcal{L}_{\rm fermion} +   \mathcal{L}_{\rm gauge}   \\
&  \sim \frac{1}{2} D_\mu S D^\mu S^* + \lambda_p  H^\dagger H S S^* + \lambda_S (S S^*)^2 + \frac{m_S^2}{2} S S^*  \nonumber \\ \nonumber
 &\quad  + i \bar{\psi} D \!\!\!\! / \psi + M_\psi \bar{\psi} \psi  + y_\psi \, S \bar{\psi} \psi^c  + h.c. \nonumber \\ 
&  \quad + \frac{1}{4}\, F^X_{\mu \nu } F_X^{\mu \nu } + \frac{\epsilon}{2}\, F^Y_{\mu \nu } F_X^{\mu \nu } + \frac{M_{Z'}^2}{2} \left( Z'_{\mu} - \partial_\mu \zeta \right)^2\nonumber\,.
\end{align}
Note that given a transformation property of $\zeta \rightarrow \zeta + \delta$ for a gauge transformation $Z'_{\mu} \rightarrow Z'_{\mu}+ \partial_\mu \delta$, the mass term for the new gauge boson is gauge invariant~\cite{Stueckelberg:1900zz}.
The gauge quantum numbers of the fermion and scalar fields are $n_\psi$ and $n_S = 2 \,n_\psi$ respectively.
The fermions are vectorlike, i.e., the left- and right-handed components of the fields carry the same quantum numbers such that the model is anomaly free.
We rotate the $U(1)$ sector in order to eliminate the mixing term $\epsilon\, F^Y_{\mu \nu } F_X^{\mu \nu }$, see App.~\ref{sec:kin-mixing} for details.
The system is then described by the mixing gauge coupling $g_\epsilon$ and the dark gauge coupling $g_D$.
The dark gauge boson covariant derivative acting on the fermion fields reads
\begin{align}
\label{eq:cov-der}
\mathcal{D}_{\mu} = \partial_\mu + i \left(g_D  n_\psi  + g_\epsilon  \,Y_f  \right) Z'_{\mu}\,,
\end{align}
where $n_\psi$ is the dark fermion gauge charge and $Y_f$ the hypercharge of a SM fermion.
 It is convenient to define $\alpha_D \equiv (n_\psi g_D )^2/ 4 \pi$ and  $\alpha_{\epsilon} \equiv (Y_f  g_\epsilon  )^2/ 4 \pi$.

This system has two relevant DM phases depending on the mass hierarchy of the involved fields. 

\subsection{Scalar dark matter}
In the case that $M_S \ll M_\psi \approx M_{Z'}$ and $\langle S \rangle = 0$, the lightest particle in the spectrum is the complex scalar field S, as discussed in Ref.~\cite{1306.4710}. Since the scalar field does not develop a vev at the low energy scale, we are left with the SM extended by a singlet scalar field and an emergent $Z_4$ symmetry ($S \rightarrow - S $ and $\psi \rightarrow i \psi $), that forbids its decay. On the other hand, in this mass hierarchy, the heavy-fermion field has to decay in order not to be overproduced. The simplest way to do so is to identify the gauge symmetry of the dark sector with the $B$-$L$ symmetry, which is anomaly free in the SM with three right-handed neutrinos.  Now the interaction with the Higgs and lepton fields $y_D\,H \bar{L} \psi_R$ and $y_D\, H \bar{L} \psi_L^c$ is allowed, SM leptons are part of the $Z_4$ symmetric subsector  ($L \rightarrow i L$) and given that $n_\psi = 1$ and the $\psi$ fermions can decay.  

However, the fermion interactions induce a decay for the DM scalar by a dimension six operator, the final states of the decay are light neutrinos. The lifetime constraints for DM imply that the fermion mass has to be above  $ M_\psi > y_D\, 10^{14} \text{ GeV}$, where $y_D$ is the coupling of the fermion decay operator.  Decay channels involving the $Z'$ are forbidden by  $B$-$L$ symmetry. 
Since the fact that the $B$-$L$ symmetry is unbroken is directly linked to DM stability in this scenario, it implies that SM neutrinos are pure Dirac particles. In reverse conclusion, this implies that if lepton number violating neutrinoless double beta decay is experimentally confirmed, this scenario would be ruled out.
 
\begin{figure}[t]
\includegraphics[width=\linewidth]{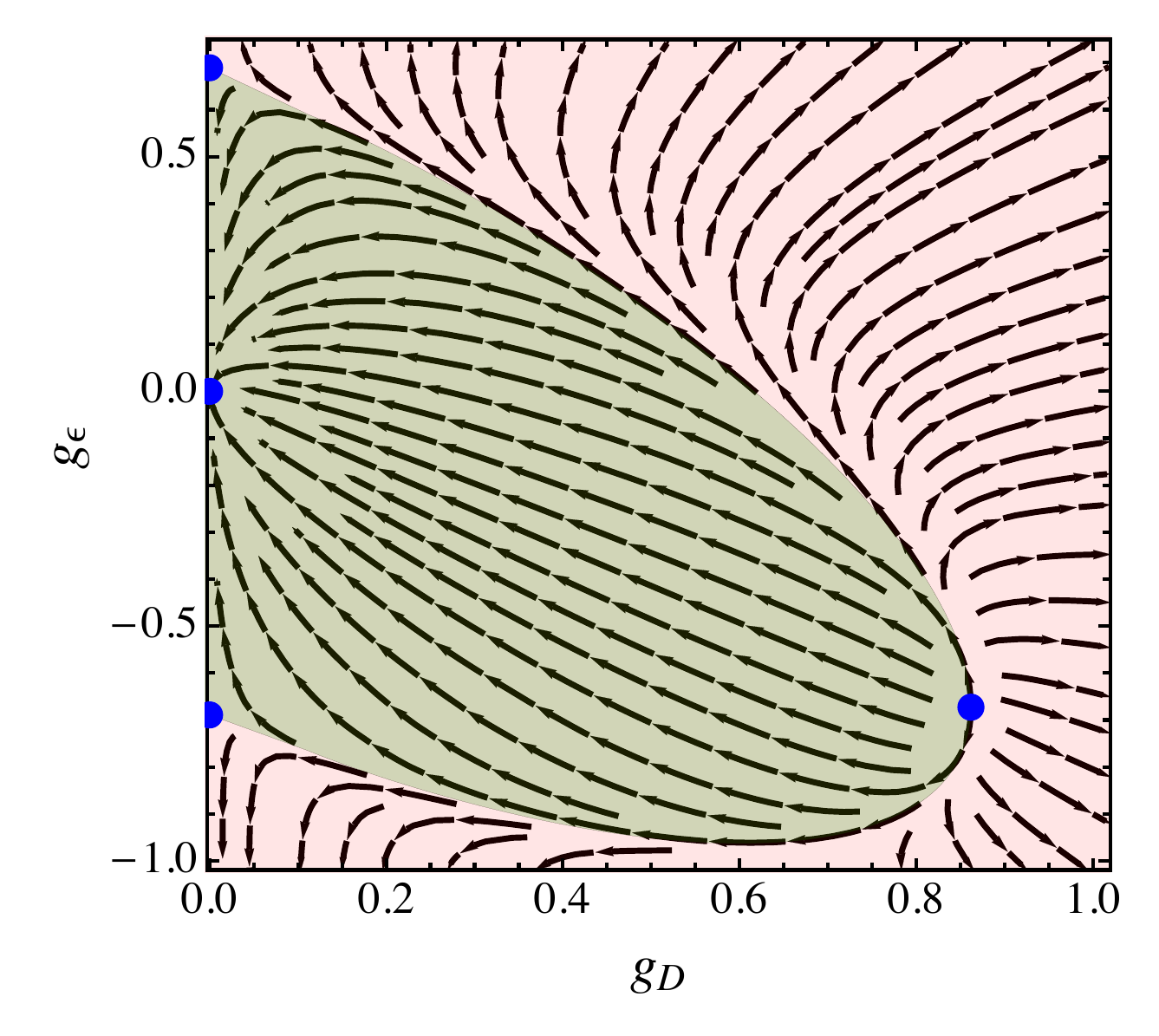}
\caption{
Favored (green) and disfavored (red) and coupling values of $g_D$ and $g_\epsilon $ at the Planck scale in the $B$-$L$ model.
The arrows indicate the RG flow towards the UV beyond the Planck scale, i.e.,
the favored coupling values flow toward the asymptotically free fixed point,
while the disfavored couplings run towards a Landau pole.
The asymptotically safe fixed points are marked with blue dots.
}
\label{fig:g-gtilde-B-L}
\end{figure}

As discussed in the previous section, asymptotic safety predicts vanishing quartic scalar couplings at the Planck scale
\begin{align}
 \lambda_p (M_\text{Pl}) = \lambda_S (M_\text{Pl}) \approx 0 \,.
\end{align}
For the $U(1)$ gauge couplings it predicts an upper bound
\begin{align}
 g_D (M_\text{Pl}) &\leq \sqrt{\frac{f_g}{\beta_{g_D,\text{1-loop}}}} \nonumber \,, \\
  g_\epsilon  (M_\text{Pl}) &\leq \sqrt{\frac{f_g}{\beta_{g_\epsilon  ,\text{1-loop}}}} \,.
 \label{eq:upbound-g}
\end{align}
The one-loop beta functions $\beta_{g_D,\text{1-loop}}$ and $\beta_{ g_\epsilon ,\text{1-loop}}$ are displayed in App.~\ref{sec:beta-functions}.
This leads to \autoref{fig:g-gtilde-B-L}, where we display the favored (green) and disfavored (red) coupling values at the Planck scale for $f_g=0.04$.
In the green region, the couplings $g_D$ and $g_\epsilon$ become asymptotically free,
while in the red region they run into a Landau pole.
The system has three interacting fixed points at $(g_D^*,g_\epsilon^*) =(0.86,-0.67)$ and $(g_D^*,g_\epsilon^*) =(0, \pm 0.69)$, where the couplings become asymptotically safe.
However, these fixed points depend on the hypercharge coupling $g$, and for $g\to0$ they turn into a line of fixed points. 

In the scalar DM phase, the system shows the following features:
\begin{itemize}
\item The gauge interaction induces the scalar portal coupling between the DM scalar and the Higgs field.
\item The fermionic contributions ensure the vacuum stability of the scalar field $S$.
\item The RG evolution of the portal coupling allows placing an upper bound on its value at the low scale.
\end{itemize}
As discussed in \autoref{sec:summary-predictivity}, we require the predicted Higgs mass to lie within $10$\,GeV around the experimentally measured value.

\begin{figure*}[t]
\includegraphics[width=.49\linewidth]{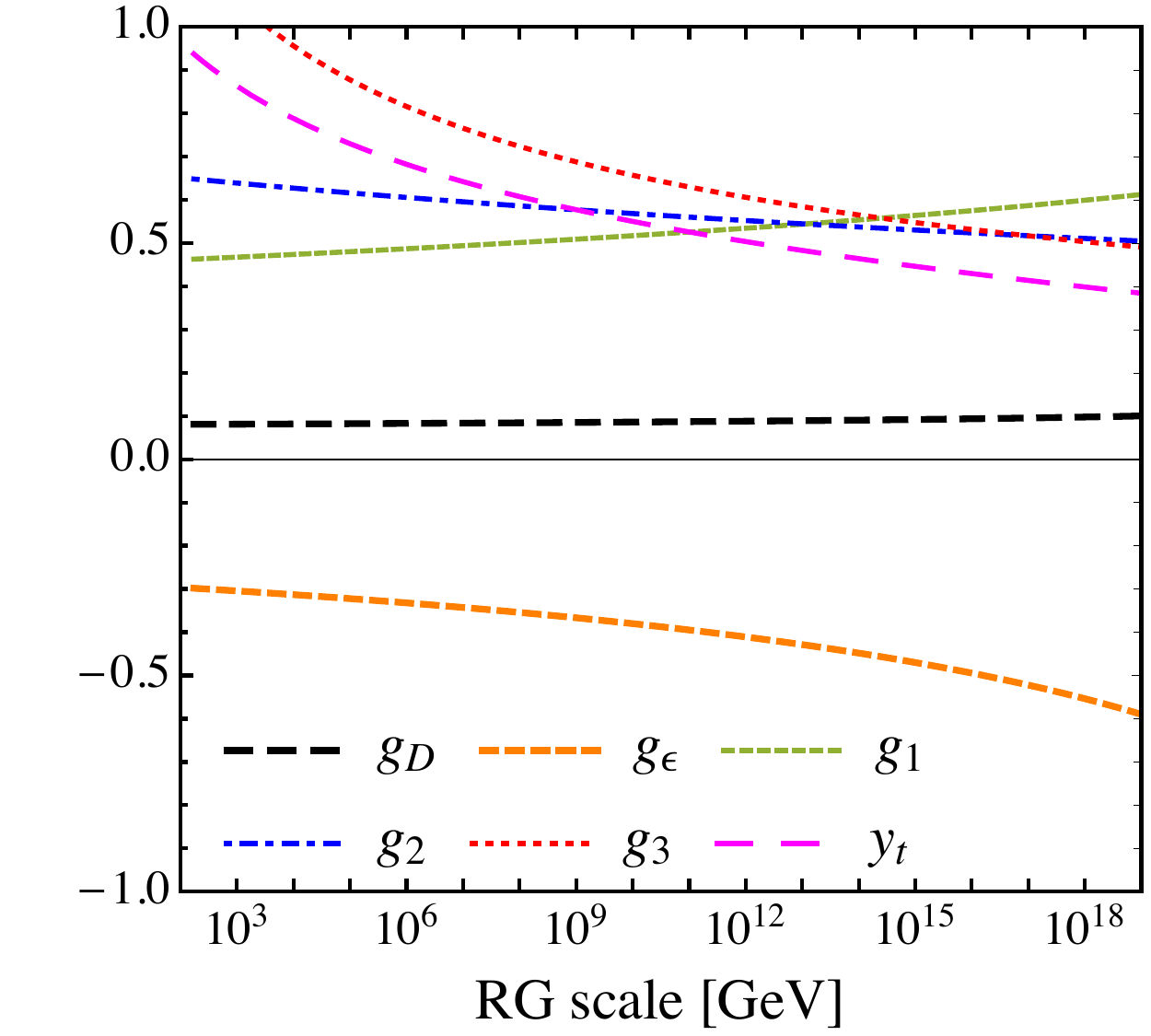} \hfill
\includegraphics[width=.49\linewidth]{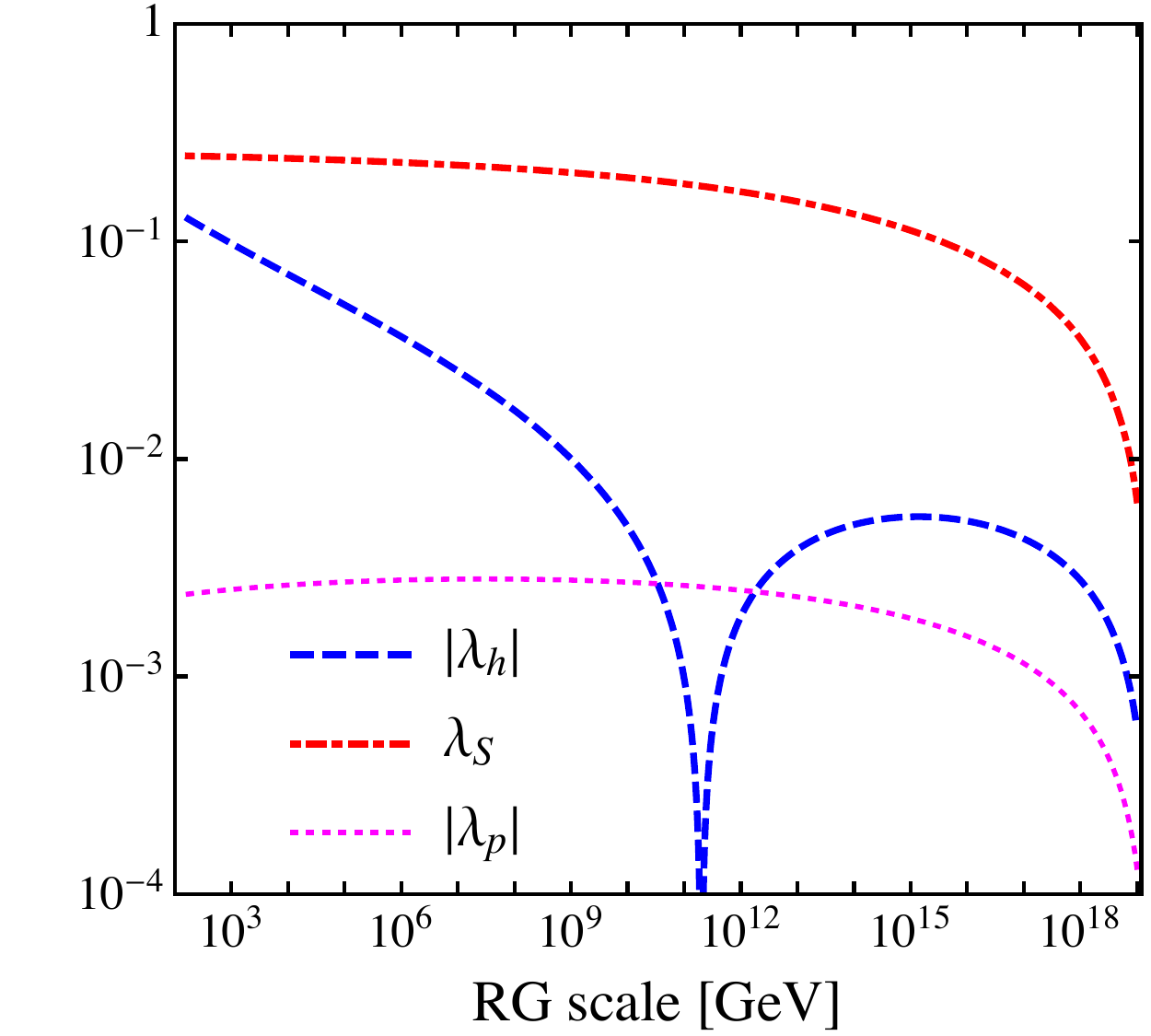}
\caption{An example of the running couplings in the $B$-$L$ model with scalar DM.
In this example we set $g_D(M_\text{Pl})=0.1$
and obtain $\lambda_p (\text{TeV})=-0.0025$ for the portal coupling.
Left: gauge and top-Yukawa couplings. Right: quartic scalar couplings.
}                    
\label{fig:example-running-couplings}
\end{figure*}

In \autoref{fig:example-running-couplings} we display an example of an RG evolution of the marginal couplings,
where all boundary conditions at the Planck scale are fulfilled and also the Higgs mass is correct.
The Higgs potential is metastable, but the lifetime of the electroweak vacuum is longer than in the SM due to the positive contributions of the new gauge interactions to the running of the quartic Higgs coupling.

By scanning the parameter space of $g_D$ and $g_\epsilon$, we find upper bounds for the portal coupling as a function of $f_g$.
We fitted the values for the maximal portal coupling and the corresponding minimal Higgs mass
\begin{align}
 \label{eq:lp-fg}
 |\lambda_{p} (\text{TeV})| &\lesssim 2.1\, f_g + 88\,f_g^2  \,,\\
 \label{eq:mh-fg}
 m_{h,\text{min}} &\approx (136 - 119 \, f_g + 2.6 \cdot 10^4 f_g^2 )\,\text{GeV} \,.
\end{align}
These bounds are derived by combining the two predictions from quantum gravity:
i) the quartic couplings at the Planck scale are approximately vanishing $\lambda_p(M_\text{Pl}) = \lambda_s (M_\text{Pl}) \approx 0$;
ii) the $U(1)$/kinetic mixing couplings $g_D$ and $g_\epsilon $, which generate the portal coupling, are limited by an upper bound at the Planck scale.
A third, important constraint is the vacuum stability of the dark scalar $S$. 
Given such boundary conditions, the RG evolution within this model does not permit larger values of the portal coupling. 

As expected from \eqref{eq:upbound-g}, the portal coupling depends roughly quadratically on the graviton contribution $f_g$.
The Higgs mass for $f_g=0$ is precisely the result for SM running at one loop.
This shifts with two-loop running to $130$\,GeV at a top pole mass of $m_t=173$\,GeV
and consequently we expect \eqref{eq:mh-fg} to be globally shifted by about 6\,GeV at two-loop order.

The fitted expressions \eqref{eq:mh-fg} illustrate an interesting mechanism in this system:
large values of the portal coupling $\lambda_p$ can only be reached in exchange for a small Higgs mass.
This can be understood from the following consideration.
A larger portal coupling can be achieved with a larger value of either $g_D$ or $g_\epsilon$.  
The running of $\lambda_h$, and thus the resulting Higgs mass, does not directly depend on $g_D$, but a large value of $g_D$ triggers a vacuum instability in the dark scalar $S$.
This constraint is tighter than the quantum gravity constraint for $f_g=0.04$ and forbids too large values of $g_D$.
Hence, one needs to increase $g_\epsilon$ to enhance the portal coupling without triggering a vacuum instability.
However, $g_\epsilon$ affects the resulting Higgs mass at leading order
and this, in summary, links a large portal coupling with a small Higgs mass.

As explained before, we restrict the Higgs mass to the interval $m_h=(125\pm 10)$\,GeV.
In that case we find 
\begin{align}
 \label{eq:lp-max}
 |\lambda_p (\text{TeV})| \leq 0.15 \,.
\end{align}
If we restrict the Higgs mass to an even tighter interval $m_h=(125\pm 1)$\,GeV,
we find a correspondingly tighter bound on the portal coupling, $|\lambda_{p}| \leq 0.1$.

\begin{figure}[t]
\includegraphics[width=\linewidth]{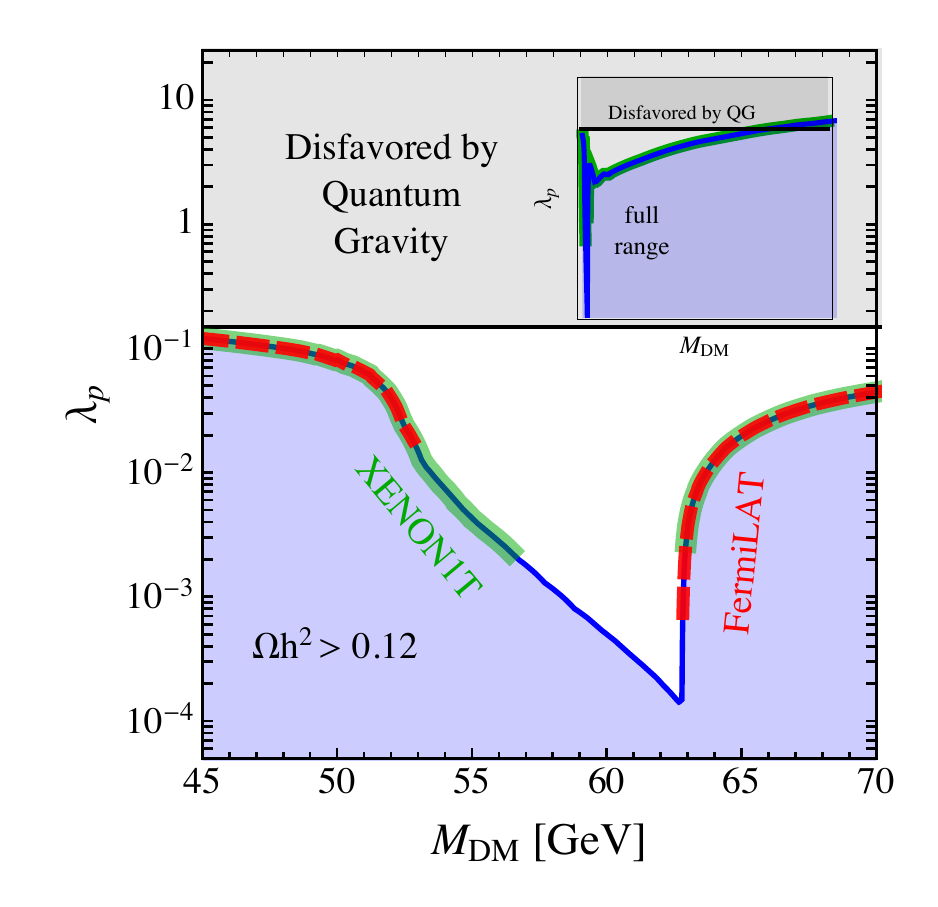}
\caption{The values of portal coupling as a function of DM mass at which the correct relic abundance is reproduced (blue line), below that line there is too much DM. Superposed are constraints from the XENON1T experiment (green line) and the FermiLAT dwarf galaxy observations (red). The grey shaded region of the parameter space is disfavored by the quantum gravity boundary conditions. The inset on the top right shows a zoom out from the resonant region to the full portal coupling parameter space.
}
\label{fig:SclalarDM}
\end{figure}

In \autoref{fig:SclalarDM}, we show that experiments exclude DM masses away from the Higgs resonance up to $\sim 2\,$TeV.
In order to reach masses above 2\,TeV, a portal coupling $|\lambda_p|\geq 0.45$ is needed,
which in turn requires $f_g\geq 0.06$ and comes at the cost of a Higgs mass of $m_h=50$\,GeV.
Such a small Higgs value is too far from its measured value and hence we either need to add new particles to change the RG flow or violate the boundary conditions at the Planck scale.
In consequence, masses above 2\,TeV are indeed not compatible with our assumptions, independent of the value of $f_g$
We emphasize that in the scalar phase of our DM model, we obtain a very stringent prediction for the DM parameters, as we will discuss in more detail in the coming section.

\subsubsection*{Relic density}
The only important DM interaction in the IR is the Higgs portal interaction. Thus, the upper bound on the portal coupling leads to a prediction for the DM mass.
Computing the $s$-channel diagrams mediated by the Higgs boson provides the required interaction cross section~\cite{1509.04282}
\begin{align}
(\sigma v_{\rm rel.})= \frac{8 \lambda_p^2 v_h^2 }{g_S \sqrt{s} } \frac{\Gamma_h(\sqrt{s} )}{(m_h^2-s)^2 + \Gamma_{h}(m_h)^2 m_h^2 }\,,
\end{align}
where $g_S$ is the number of scalar degrees of freedom, $m_h, v_h$ the Higgs boson mass and vev and $\Gamma_h(\sqrt{s})$ the momentum dependent Higgs decay width.

We apply the boundary layer method~\cite{1203.1822} to obtain an asymptotic solution for the Boltzmann equation and determine the relic density. Note that since we are interested in values close to the Higgs resonance, the full thermal average of the cross sections has to be performed~\cite{Gondolo:1990dk}
\begin{align}
\langle \sigma v_{\rm rel.}\rangle = \int_{4 M_S^2}^\infty \!\! \frac{(\sigma v_{\rm rel.})\,s \sqrt{s- 4 M_S^2}\, K_1(\sqrt{s}/T)}{16\, T\, M_S^4\, K_2^2(M_S/T)} \mathrm ds \,.
\end{align}
Here, $K_1$ and $K_2$ are modified Bessel functions of the first and second kind and $M_S^2 = m_S^2 + \lambda_p v_h^2$ is the DM mass.

In \autoref{fig:SclalarDM}, we show the values of the portal coupling as a function of the DM mass for which the cosmological relic density constraint is satisfied.
The upper bound from quantum gravity is indicated by the horizontal line.
As displayed, the relic density constraint is only satisfied if the DM mass is in the vicinity of the Higgs resonance 45\,GeV$ <  \MDM < 500$\,GeV.
In that case, the annihilation is resonantly enhanced and, despite the small coupling, enough DM annihilates away to reproduce its cosmological abundance. Current experimental constraints from XENON1T~\cite{1805.12562} and the FermiLAT~\cite{1601.06590} experiment restrict the allowed parameter space even further around the Higgs resonance, leaving the viable DM mass in a narrow range 56\,GeV$ <  \MDM < 63$\,GeV. We find that in the scalar DM phase the model is highly predictive. 

The FermiLAT bounds are derived from the limits on the annihilation cross section to b-quarks, since this is the dominant channel in the considered mass range. For the XENON1T limits, we used the DM-nucleon cross section given by 
\begin{align}
\sigma_{\rm SI} = \frac{\lambda_p^2 f_N^2 m_N^2 \mu^2}{\pi m_h^4\,M_S^2}\,,
\end{align}
 where $\mu$ is the reduced mass of the DM-nucleon system and $f_N$ the effective Higgs-nucleon coupling, with the best current value of $f_N \approx 0.308 \pm 0.018$~\cite{1708.02245}.

\subsection{Fermionic dark matter}
In the case that $M_\psi  \ll  M_S$ and $M_{Z'}  \ll M_S$, the lightest stable particle in the dark sector is the new fermion.
For non-zero values of the quantum number $n_\psi$, the coupling to the lepton doublet through the Higgs field ($L \,H \psi$)  is forbidden and the new fermion does not decay to SM particles. The stability is thus accidental and related to the quantum number choice of the fermion. The mass of the gauge boson of the dark sector force has a contribution from the Stueckelberg term in \eqref{eqn:dmmodel}. Furthermore, the scalar $S$ can get a vev, in which case the gauge boson has two mass contributions. Both the scalar and vector fields are unstable due to the presence of Yukawa and gauge interaction in the sector.  As we discuss shortly, the heavy scalar does not play a role for DM phenomenology in this mass hierarchy. 

\begin{figure}[t]
\includegraphics[width=\linewidth]{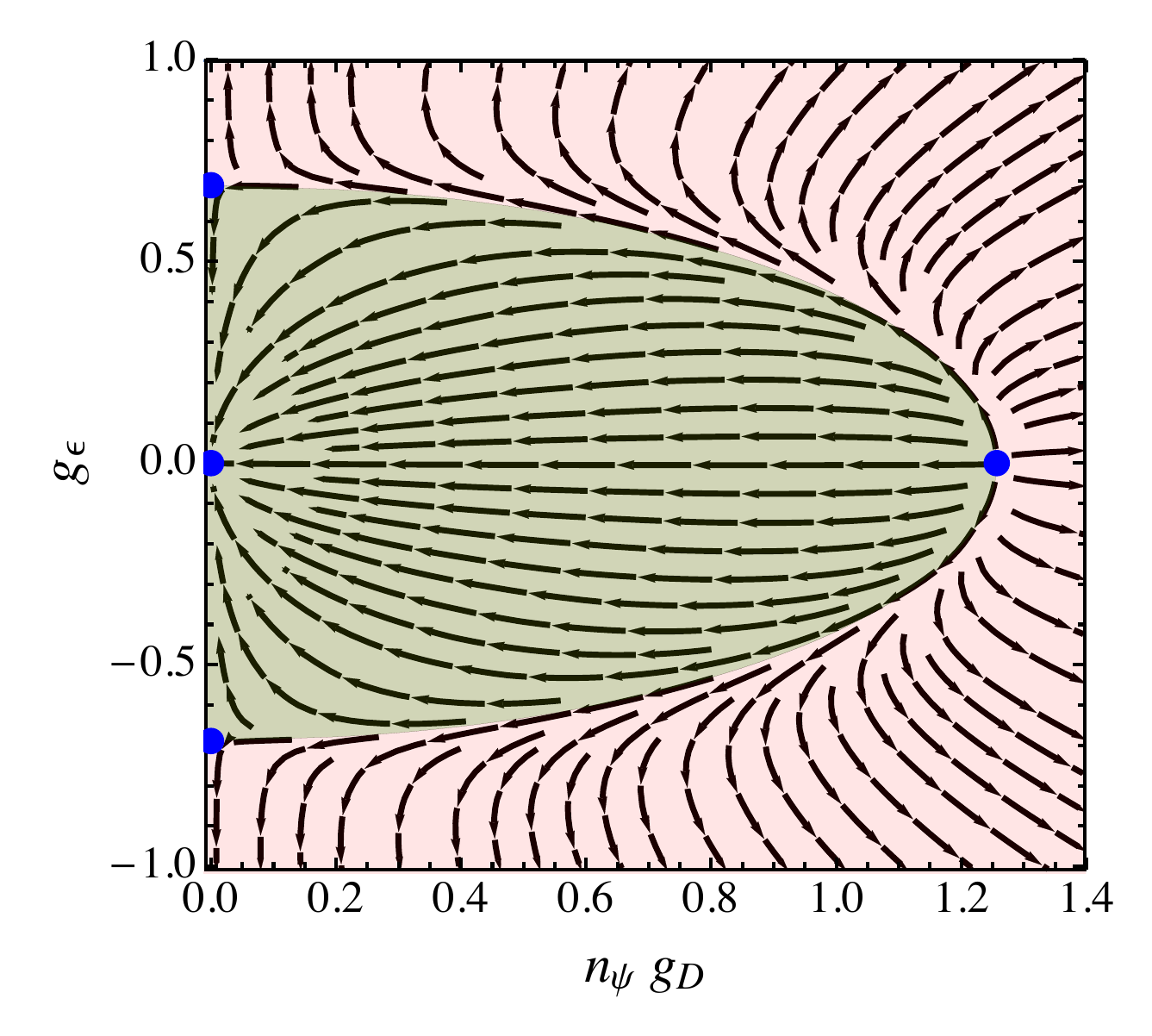}
\caption{
Favored (green) and disfavored (red) and coupling values of $n_\psi\,g_D$ and $g_\epsilon $ at the Planck scale in the $U(1)_X$ model.
The arrows indicate the RG flow towards the UV beyond the Planck scale, i.e.,
the favored coupling values flow toward the asymptotically free fixed point,
while the disfavored couplings run towards a Landau pole.
The asymptotically safe fixed points are marked with blue dots.
}
\label{fig:g-gtilde-plane}
\end{figure}

The requirement of an asymptotically safe theory including quantum gravity limits the values of the new gauge coupling and the kinetic mixing.
As shown in \eqref{eq:upbound-g} this upper bound depends on the one-loop coefficient of the gauge beta function, which in turn depends on the quantum number of the fermion.
This is a crucial ingredient for the predictivity of the model in this mass hierarchy:
If the fermion has a small quantum number, then the gauge coupling can have a large value at the Planck scale.
On the other hand, a large quantum number restricts the gauge coupling to be small at the Planck scale.
For the relic density, only the combination of the quantum number with the gauge coupling enters and this mechanism keeps this roughly constant.

In \autoref{fig:g-gtilde-plane}, we show the range for the couplings $g_D$ and $g_\epsilon $ favored by asymptotically safe quantum gravity for $f_g=0.04$.
In the green region, the couplings $g_D$ and $g_\epsilon$ become asymptotically free,
while in the red region they run into a Landau pole.
This system has three interacting fixed points at $(g_D^*,g_\epsilon^*) = (1.26,0)$ and $(g_D^*,g_\epsilon^*) = (0,\pm  0.69)$.
Again, these fixed points depend on the hypercharge coupling $g$, and for $g\to0$ they turn into a line of fixed points. 
If we consider each coupling separately, we obtain the bounds
\begin{align}
n_\psi\,g_D (M_\text{Pl}) &\leq 1.26 \,, \nonumber \\
\left| g_\epsilon  (M_\text{Pl}) \right| &\leq 0.69 \,.
\end{align}
The lowest relic density, and consequently the largest mass is, however, obtained with the largest product of the two couplings as detailed in the next section.
Consequently, this translates into the following upper bounds on the interaction parameters
\begin{align}
\label{eq:prediction-g-b}
n_\psi g_D (M_\text{Pl}) g_\epsilon  (M_\text{Pl}) \leq 0.43 \,.
\end{align}
Since the gauge boson provides the link between the DM and the SM sector, those couplings are the most important ones for our considerations. At low energy scales, the relatively mild RG running of those couplings leads to the following maximally attainable values $n_\psi g_D (\text{TeV}) \leq 0.51$ and $ g_\epsilon  (\text{TeV}) \leq 0.16$ or equivalently 
\begin{align}
\label{eq:prediction-alpha}
\alpha_D(\text{TeV})&\equiv \frac{\left(n_\psi g_D(\text{TeV})\right)^2}{4 \pi} \leq 0.021\,,\notag \\
\alpha_\epsilon(\text{TeV})&\equiv \frac{\left(Y_f  g_\epsilon(\text{TeV})\right)^2}{4 \pi} \leq 2\cdot 10^{-3}\,.
\end{align}
In this scenario the Higgs mass is only affected at the percent level and does not provide additional constraints on the model parameters.

\subsubsection*{Relic density}
\begin{figure}[t]
\includegraphics[width=\linewidth]{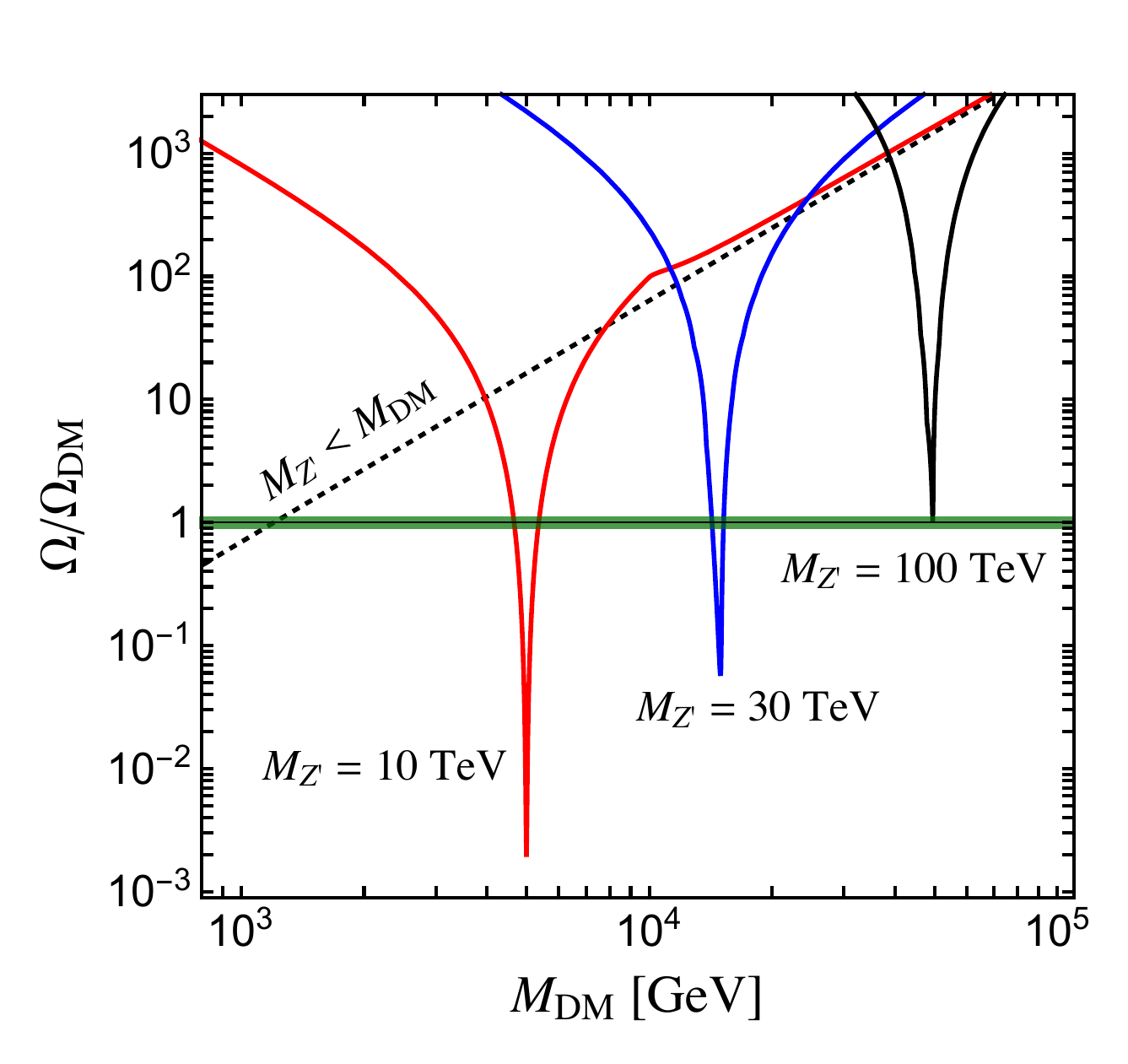}
\caption{The relic density for different fermionic DM masses given three mediator mass choices assuming maximal couplings that are still compatible with asymptotically safe quantum gravity. The maximal DM mass at which DM is not overproduced is 50\,TeV, in maximal resonance with a 100\,TeV force mediator. The gauge couplings are such that their product maximizes the quantum-gravity bound, see \eqref{eq:prediction-alpha}. }
\label{fig:FermionDM}
\end{figure}
If $M_{Z'} > M_\psi$, then the relic density is set by an $s$-channel process where the new gauge boson is exchanged. The coupling to the SM particles is controlled by the kinetic mixing between the dark sector gauge boson and the hypercharge gauge boson. The cross section for the annihilation into a pair of SM fermions is given by 
\begin{align}
(\sigma v_{\rm rel.}) = \frac{4 \pi  \alpha_D \alpha_{\epsilon } }{3 s } \frac{\sqrt{\frac{s - 4 m_{f}^2}{M_{\psi}^2}} \left( 2 m_{f}^2+s \right) \left(2 M_{\psi}^2+s \right)}{ \left( M_{Z'}^2- s\right)^2 + \Gamma_{X}^2 M_{Z'}^2}\, ,
\end{align}
For the total cross section, we sum over all kinematically accessible final states. As in the scalar DM case, the thermal average is performed following~\cite{Gondolo:1990dk}, since we are dealing with processes close to the resonance. 

An additional annihilation channel is possible, if the dark sector scalar gets a vev and a mixing with the Higgs boson is induced 
\begin{align}
(\sigma v_{\rm rel.}) =  \frac{y_\psi^2\,\sin^2(\theta) \left(s - 4 M_{\chi}^2 \right) \,\Gamma_h(\sqrt{s})}{ \sqrt{s}\, \left( \left( M_{S}^2- s\right)^2 + \Gamma_{S}^2 M_{S}^2 \right) }\,.
\end{align}
However, this cross section is velocity suppressed, as it is a $p$-wave process. Additionally, in the mass hierarchy regime $M_S \gg m_h$, given the quantum gravity bound on the portal coupling, the mixing angle $\sin(\theta) = \lambda_p v_h v_S/M_S^2 \approx  \lambda_p v_h/M_S$ is bound to be below $\mathcal{O}(10^{-4})$. The Yukawa coupling, as discussed in \autoref{sec:yukawa}, is also bounded by the quantum gravity boundary condition to be $y_\psi \lesssim \mathcal{O}(1)$. We therefore assume, that even if the dark scalar is of similar mass as the dark gauge boson, the cross section of the Yukawa channel is subdominant.

If $M_\psi > M_{Z'}$, then the dominant process is a $t$-channel interaction leading to $\psi + \psi \rightarrow Z'_\mu + Z'_\mu$ and the gauge bosons $Z'_\mu$ decay into SM particles due to kinetic mixing.  The dominant $s$-wave contribution to the cross section is given by
\begin{align}
\langle \sigma v_{\rm rel.} \rangle \approx \frac{ \pi \alpha_D^2}{M_\psi^2}  \left( 1 - \frac{M_{Z'}^2}{M_\psi^2}\right)^{3/2} \left( 1 - \frac{M_{Z'}^2}{2 M_\psi^2}\right)^{-2}\,.
\end{align}
Since $\alpha_\epsilon$ can be very small in this scenario, we used the global maximum, as a limit for the dark gauge coupling $\alpha_D < 0.03$. Note that since $\alpha_D$ is restricted to relatively small coupling values, the Sommerfeld enhancement, which is possible in this regime, if $M_{Z'} \ll M_\psi$, is only marginal during the freeze-out, but might become relevant at low velocities today. 

In \autoref{fig:FermionDM}, the relic abundance is shown as a function of the mass of the fermionic DM candidate $\psi$. In the heavy mediator regime, i.e., $M_{Z'} > M_\psi$, the measured DM relic density can only be obtained close to the resonant regime. Analogously to Refs.~\cite{1409.8165, 1506.05107, 1803.07462, 1810.06646}, an upper bound can be obtained on the DM mass in this case, by assuming maximal on-resonance annihilation. Our upper bound is more robust than previously assumed as the DM gauge charge enters the effective coupling $\alpha_D$ and is, thus, constrained by the requirement of asymptotic safety as well. In contrast, to~\cite{1506.05107}, where the upper bound was a function of the free DM gauge charge, in our case, the DM mass is generically bounded to be $M_\psi \lesssim 50$\,TeV. 

One may ask the question of whether reducing the coupling value of $\alpha_\epsilon$, which leads to smaller gauge boson decay width, if one may reach a more extreme resonance and, thus, a larger DM mass in this finely tuned regime. However, due to thermal effects on the velocity averaged cross section a further reduction in the decay width does not compensate for the scaling of the cross section with $\alpha_\epsilon$. Thus, in this manner no larger DM masses can be reached. 

The derived upper bound on the mass holds also in a mixed scenario, where scalar and fermion fields are stable. Assuming that part of the DM is made up of stable scalar particles $S$, as discussed in the previous section, less DM can be made of heavy fermions $\psi$. Since its relic density scales as $\Omega_\psi \propto M_\psi^2$, the upper mass bound only gets tighter in such a multi-component DM scenario.  

In the case that $M_{Z'} < M_\psi$, the upper mass bound is even more severe and the maximal attainable DM mass is $M_\psi < 2$\,TeV. For this scenario to be phenomenologically viable, however, the coupling related to the kinetic mixing parameter $\alpha_\epsilon$ has to be very small leading to relatively long life-times for the $Z'_\mu$, see for example Ref.~\cite{0711.4866}. Since the annihilation process through the vector mediator is an $s$-wave process, severe bounds from the CMB rule out DM masses below $\mathcal{O}(10)$\,GeV~\cite{1909.08632} and a combined analysis of indirect detection experiments leads to $\mathcal{O}(20)$\,GeV~\cite{1805.10305}.  

Finally, there is one configuration, which can also lead to a very light (sub-GeV) DM scenario. This is possible if the scalar $S$ decays through mixing with the Higgs boson and serves as the mediator to the SM. This scenario is unconstrained by CMB observations since the annihilation proceeds through a $p$-wave process and is strongly suppressed at late times. The mass hierarchy $M_S > M_\psi$ is excluded in the light DM scenario~\cite{1512.04119}.  However, in the opposite regime  $M_S < M_\psi$ DM as light as  $\mathcal{O}(10)\, \rm MeV$ can be thermally produced. 

\subsection{Experimental Searches}
\label{sec:experimets}

In this section, we briefly summarize the viable DM scenarios in the gauge-Yukawa dark sector embedded in asymptotically safe quantum gravity and discuss their experimental accessibility. 

\begin{enumerate}
\item Scalar DM coupled via the Higgs portal in the resonant configuration, with $\MDM \approx m_h/2$. The portal coupling in this regime can be as small as $\lambda_p \approx 10^{-4}$, however, even with such small couplings the spin-independent nucleon cross section is of the order of $\sigma_{\rm SI} \approx 5 \cdot 10^{-49} \, \text{cm}^2$. Despite being small, this cross section is above the neutrino floor and, thus, testable by large volume liquid noble gas detectors, such as DARWIN~\cite{1606.07001}. On the other hand, searches for the annihilation signal in space will be able to probe this parameter region as well once an improvement of about two orders of magnitude in sensitivity takes place. Intriguingly, there are astrophysical observations, which might be explained by DM annihilation in that mass range~\cite{1510.07562,1904.08430,2001.08749}. Given the branching ratios of the Higgs boson, this scenario predicts a gamma line with $E_\gamma \approx 60 \, \rm GeV$, corresponding to an annihilation cross section of $\langle \sigma v_{\rm rel.} \rangle_{\gamma \gamma} \approx 10^{-29} \,\text{cm}^3/\text{s}$.

Also searches for antiparticles in cosmic rays can provide a confirmation of this scenario. A crucial observable is the ratio of anti-helium to anti-deuterons $R_3 = ^3\!\!\overline{\text{He}}/\bar{d}$ and  $R_4 = ^4\!\!\overline{\text{He}}/\bar{d}$. In this DM scenario, with the dominant annihilation mode being $\text{DM} \text{DM} \rightarrow b \bar{b}$, those ratios are expected to be $R_3 \approx 3 \times10^{-2}$ and $R_4 \approx 10^{-5}$, exceeding those expected from astrophysical sources $R_3 \approx 10^{-2}$ and $R_4 \approx 10^{-8}$ , see Ref.~\cite{2001.08749} for a detailed discussion. Overall, the flux of heavier antiparticles from DM annihilation in this scenario is expected to be at least an order of magnitude larger than the flux from astrophysical sources and within reach of the \textit{AMS-02}  experiment~\cite{Aguilar:2016kjl}.

\item Fermion DM coupled via the $Z'$ portal.  Here, two mass hierarchy regimes can lead to distinctly different phenomenologies. The first is the light mediator regime with $\MDM > M_{Z'}$ with allowed DM masses between $\mathcal{O}(10)\,\text{GeV} < \MDM < 2 \,\text{TeV}$. Here, due to a potentially large life-time of the mediator, the annihilation signal from Dwarf galaxies and the Galactic center can be significantly softened and challenging to detect. However, searches that measure the total energy injections, such as CMB observations~\cite{0906.1197} or radio wave observations of the early Universe~\cite{1803.03629}, can explore this scenario efficiently.  Since in this case, we do not have a prediction for the coupling strength $\alpha_\epsilon$ from the freeze-out condition, the direct detection signal can potentially be very small.

The second is the heavy mediator mass regime, where the correct relic abundance can only be reproduced close to the $s$-channel resonance, i.e., $\MDM \approx 2 M_{Z'}$. The spin independent cross section is given by $\sigma_{\rm SI} \approx 1.8 \cdot10^{-38} \alpha_D \, \alpha_\epsilon (\text{TeV}/\MDM)^2 \text{ cm}^2 \approx  1.5 \cdot10^{-42}  (\text{TeV}/\MDM)^2 \text{ cm}^2 $. Thus, in the resonant scenario XENON1T excludes DM masses below $2.5$\,TeV and the cross section remains above the neutrino floor up to a DM mass of $9$\,TeV.
However, in the heavy mediator mass regime DM masses can be as large as $\MDM \sim 50$\,TeV and the most promising search strategy seems the search for the annihilation signals with future experiments, such as the Cherenkov Telescope Array, CTA~\cite{1008.3703}.

A further bound is provided by hidden $U(1)$ gauge boson searches at the LHC \cite{1212.3620}.
Here, the kinetic-mixing coupling $\epsilon=-g_\epsilon/\sqrt{g_Y^2+g_\epsilon^2}$ is constrained as a function of the mediator mass.
For $M_{Z'}\leq2.5$\,TeV\,(1\,TeV) the bound is $|g_\epsilon|\leq 0.22$\,(0.015), while there is no bound for larger $M_{Z'}$.
However, for these mediator masses, we can choose $g_\epsilon$ small enough and still reach the correct DM relic abundance.
The same holds in the light-mediator regime where $g_\epsilon$ can be chosen very small.
Hence, colliders become insensitive to this scenario. On the other hand, observations of the sun can investigate parts of the parameter space with long-lived $Z'$~\cite{1808.05624}.

\item Fermion DM coupled though the scalar-Higgs portal. In this scenario, masses of the order of a few GeV are already excluded by direct detection searches, but the mass window between $\mathcal{O}(10)\,\text{MeV} < \MDM < \mathcal{O}(5)\,\text{GeV}$ remains open. The relic density can be set in this mass regime if the scalar mediator mass is lower than the DM fermion mass~\cite{1909.08632}. Annihilation signals are strongly suppressed in this scenario, and the best way to explore its parameter space is direct searches with lower detection thresholds, see Refs.~\cite{1708.08929,1810.06283}.  
\end{enumerate}

\section{Discussion of uncertainties}
\label{sec:discussion}
Given our results, it is important to discuss the uncertainties and caveats of the approach.

Our most basic hypothesis is that we consider only models that are asymptotically safe after the inclusion of quantum gravity.
Quantum gravity is treated here as a fundamental non-perturbative QFT.
In the case of a different embedding of the SM such as string theory, our results only hold if the asymptotically safe fixed point serves as an attractor of the RG flows~\cite{1907.07894}.

While there are many hints for the existence of the asymptotically safe fixed point,
quantitative control over the fixed point is not yet achieved. 
This is related to the enormous amount of tensor structures in gravity 
and the scheme dependence of graviton contributions.
Hence, using a certain value for, e.g., the coefficient $f_g$ has to be taken with caution.
Thus, we have used a rather large range for $f_g$, trying to account for the large uncertainty.
We emphasize, however, that in the case of the scalar DM scenario, coupled to the SM through the Higgs portal, our conclusions are essentially independent of the exact value of $f_g$.

In our approach quantum gravity provides Planck-scale boundary conditions and thus it would be interesting to study more general guiding principles, which can provide boundary conditions at high energies. For example, scale invariance or conformal symmetry could be such concepts and questions such as the value of the Higgs mass~\cite{1112.2415} and the viability of SM extensions has been investigated under this assumption, see for example~\cite{1310.4423,1405.6204,1503.03066,1603.03603}.

Additionally, to the discussion of the generality of our approach to estimating the effects of quantum gravity, we can also raise the question of how general our DM framework is. As argued in~\cite{1506.03116}, a simplified DM model can be only a part of a UV complete sector, but efficiently capture the relevant information for the DM production detection. However, in our approach, we go further and raise the question of what a dark sector can look like with fields at a much lower scale than the Planck scale. In particular, this implies that there should be no Landau poles between the low energy scale and the Planck scale, given that we consider an Abelian SM extension. We find that RG stability favors a gauge-Yukawa theory. In that sense, our dark sector construction is indeed rather general. 

A logical extension of this scenario would be the introduction of a DM candidate charged under a non-Abelian interaction. This interaction could be either of a SM force, see for example Refs.~\cite{1801.01135,1904.11503} or a new non-Abelian interaction, see Refs.~\cite{1503.08749,1707.05380,1811.06975}. Relevant constraints can also be obtained in these scenarios, in particular, if the non-Abelian gauge coupling is not asymptotically free by itself. However, we defer the investigation of non-Abelian dark sectors embedded in asymptotically safe quantum gravity to future work. 

\section{Conclusions}
\label{sec:summary}
In this work, we have investigated the interplay of dark matter and asymptotically safe quantum gravity.
The assumption that quantum gravity and all matter couplings are asymptotically safe or free leads to boundary conditions at the Planck scale.
These boundary conditions, in turn, lead to the mass constraints of the dark matter candidate.

We applied this formalism to two minimal dark matter scenarios.
The requirement that the dark matter candidate is stable or long-lived and has a portal coupling to the SM, as well as stable RG trajectories up to the Planck scale naturally led us to a gauge-Yukawa theory.
We introduced a new $U(1)_X$ gauge group, a scalar via the Higgs portal and dark vectorlike fermions.
Depending on the mass hierarchy, either the scalar or the fermion is the dark matter candidate.

For the scalar dark matter, we identified the new gauge group with $B$-$L$.
The model is predictive because quantum gravity demands a vanishing portal coupling at the Planck scale and sets an upper bound on the new gauge interactions.
As a consequence, only small portal couplings are reachable in the IR.
We find that the model is highly predictive in this scenario.
Due to the experimental constraints of XENON1T and FermiLAT, 
the allowed mass range for the dark matter candidate is 56\,GeV$ <  \MDM < 63$\,GeV.

For fermionic dark matter, the predictive power of the model relies on an interesting mechanism.
The boundary conditions from quantum gravity on the new gauge couplings depend on the quantum number of the dark fermion.
However, also the production rate depends on this number and, remarkably, these dependencies cancel each other.
This makes this model highly predictive.
If the mediator gauge boson is heavier than the dark fermion, the mass bound is given by $M_\psi \leq 50$\,TeV.
If the mediator gauge boson is lighter, then the bound is even tighter, $M_\psi \leq 2$\,TeV.

\vspace{.3cm}
\noindent {\bf Acknowledgements}
We thank John Beacom, Christopher Cappiello, Astrid Eichhorn, Kevin Ingles, Pavel Filevies Perez, Stuart Raby, Frank Saueressig, Masatoshi Yamada and Bei Zhou for very helpful comments on the manuscript. 
This work is partially supported by the Danish National Research Foundation grant DNRF:90. JS acknowledges support by the Alexander von Humboldt foundation. 

\appendix

\section{Kinetic mixing}
\label{sec:kin-mixing}
An important ingredient for the scalar DM model is the kinetic mixing between the two $U(1)$ gauge sectors.
This allows us to generate the portal coupling $\lambda_p$ without the Higgs boson being charged under the new gauge group $U(1)_X$.
We present here the computational details and follow closely the discussions in Refs.~\cite{1510.02379,1608.07271}.

The Lagrangian including kinetic mixing is given by
\begin{align}
\label{eq:lagrangian-kin-mix}
\mathcal{L}\sim \frac{1}{4}\, F^Y_{\mu \nu } F_Y^{\mu \nu } +\frac{1}{4}\, F^X_{\mu \nu } F_X^{\mu \nu } + \frac{\epsilon}{2}\, F^Y_{\mu \nu } F_X^{\mu \nu }\,.
\end{align}
The term $F^Y F^X$ can be eliminated by a rotation and a rescaling of the gauge fields. The transformation
\begin{align}
\begin{pmatrix} A_\mu^1 \\ A_\mu^2 \end{pmatrix}
 = \frac{1}{\sqrt{2}}
 \begin{pmatrix}
  \frac{1}{\sqrt{1-\epsilon}} & -\frac{1}{\sqrt{1+\epsilon}} \\
  \frac{1}{\sqrt{1-\epsilon}} & \frac{1}{\sqrt{1+\epsilon}}
 \end{pmatrix}
 \begin{pmatrix} B_\mu^1 \\ B_\mu^2 \end{pmatrix},
\end{align}
brings \eqref{eq:lagrangian-kin-mix} in diagonal shape.
The price to pay is that the covariant derivative is now non-diagonal
\begin{align}
 \mathcal{D}^\mu = \partial^\mu &+ i (q_Y g_{11} + q_X g_{21}) B_1^\mu \notag \\ &+ i (q_Y g_{12} + q_X g_{22}) B_2^\mu  \,,
\end{align}
where $q_Y$ and $q_X$ are the charges under the respective gauge group and the couplings $g_{ij}$ are given by
\begin{align}
\label{g-relation-1}
\frac{1}{\sqrt{2}}\begin{pmatrix} g_1 & 0 \\ 0 & g_2 \end{pmatrix}
\begin{pmatrix}
  \frac{1}{\sqrt{1-\epsilon}} & -\frac{1}{\sqrt{1+\epsilon}} \\
  \frac{1}{\sqrt{1-\epsilon}} & \frac{1}{\sqrt{1+\epsilon}}
 \end{pmatrix}
= 
\begin{pmatrix} g_{11} & g_{12} \\ g_{21} & g_{22} \end{pmatrix}.
\end{align}
The computation of the beta functions is most convenient in this basis.
We obtain the beta functions for $g_{11}$, $g_{12}$, $g_{22}$, and $g_{22}$ from \textit{PyR@TE 2} \cite{Lyonnet:2013dna,Lyonnet:2016xiz}.
However, these couplings are not independent, which is visible in \eqref{g-relation-1}.
They fulfil the relation
\begin{align}
\label{eq:relation}
g_{11} \, g_{22} = - g_{12} \, g_{21} \,.
\end{align}
For the physics, it is more convenient to parameterize the couplings in terms of the three independent couplings $g_Y$, $g_\epsilon$, and $g_D$.
This is achieved by the rotation
\begin{align}
\label{eq:g-relation}
\begin{pmatrix}
g_Y & 0 \\
g_\epsilon & g_D
\end{pmatrix}
=
\begin{pmatrix}
g_{11} & g_{12} \\
g_{21} & g_{22}
\end{pmatrix} O_R^T \,,
\end{align}
where
\begin{align}
O_R=
\frac{1}{\sqrt{g_{22}^2+g_{21}^2}}
\begin{pmatrix}
g_{22} & -g_{21} \\
g_{21} & g_{22}
\end{pmatrix}.
\end{align}
With this we arrive at the covariant derivative given in \eqref{eq:cov-der}.
This rotation also transforms the gauge field into their standard form
\begin{align}
\begin{pmatrix} B_\mu \\ Z'_\mu \end{pmatrix}
 = O_R \begin{pmatrix} B_\mu^1 \\ B_\mu^2 \end{pmatrix}.
\end{align}
In order to obtain the beta functions for $g_Y$, $g_\epsilon$, and $g_D$
we take a scale derivative of \eqref{eq:g-relation}.
We plug in the computed beta functions of $g_{11}$, $g_{12}$, $g_{22}$, and $g_{22}$.
Finally, \eqref{eq:g-relation} together with \eqref{eq:relation} allows us to express $g_{11}$, $g_{12}$, $g_{22}$, and $g_{22}$
in terms of $g_Y$, $g_\epsilon$, and $g_D$.
This yields the beta functions for $g_Y$, $g_\epsilon$, and $g_D$ displayed in the next appendix.

\begin{widetext}
\section{Beta functions}
\label{sec:beta-functions}
We used the package \textit{PyR@TE 2} \cite{Lyonnet:2013dna,Lyonnet:2016xiz} for the derivation of the beta functions.
In all considered models the beta functions for the SM gauge couplings remain unchanged at one-loop order (up to the gravity contributions).
They are given by
\begin{align}
(4 \pi)^2 \beta_{g} &=  - \tilde f_g g + \frac{41}{6} g^3 \,,
&
(4 \pi)^2 \beta_{g_2} &=  - \tilde  f_g g_2-\frac{19}{6} g_2^3 \,,
&
(4 \pi)^2 \beta_{g_3} &=  - \tilde f_g g_3-7 g_3^3  \,,
\end{align}
where we introduced $\tilde f_i = (4 \pi)^2 f_i$.

\subsection{\texorpdfstring{$B$-$L$}{B-L} model}
The beta functions in the $B$-$L$ model are given by
\begin{align}
(4 \pi)^2 \beta_{g_D} &= -\tilde f_g g_D+(12 + \frac23 N_f) g_D^3+\frac{32}{3} g_D^2 g_\epsilon + \frac{41}{6} g_D g_\epsilon ^2 \,,\\
(4 \pi)^2 \beta_{g_\epsilon } &=  -\tilde f_g g_\epsilon +\frac{32}{3} g^2 g_D + \frac{41}{3} g^2 g_\epsilon + (12 + \frac23 N_f) g_D^2 g_\epsilon +\frac{32}{3} g_D g_\epsilon ^2 +\frac{41}{6} g_\epsilon ^3 \,,\\
(4 \pi)^2 \beta_{\lambda_h} &=  \tilde f_\lambda \lambda_h+\frac{3}{8} g^4 + \frac{3}{4} g^2 g_2^2 + \frac{3}{4} g^2 g_\epsilon ^2-3 g^2 \lambda_h+\frac{9}{8} g_2^4 + \frac{3}{4} g_2^2 g_\epsilon ^2 \notag \\
   &\quad -9 g_2^2 \lambda_h+\frac{3}{8} g_\epsilon ^4-3 g_\epsilon ^2 \lambda_h+24 \lambda_h^2+\lambda_p^2-6 y_t^4+12 \lambda_h y_t \,,\\
(4 \pi)^2 \beta_{\lambda_s} &=   \tilde f_\lambda \lambda_s+ 96 g_D^4-48 g_D^2 \lambda_s+20 \lambda_s^2+2 \lambda_p^2  - 16 N_f y_\psi^4 + 8 N_f y_\psi^2 \lambda_s  \,,\\
(4 \pi)^2 \beta_{\lambda_p} &= \tilde f_\lambda \lambda_p-\frac{3}{2} g^2 \lambda_p+12 g_D^2 g_\epsilon ^2 - 24 g_D^2 \lambda_p-\frac{9}{2} g_2^2 \lambda_p - \frac{3}{2} g_\epsilon ^2 \lambda_p +12 \lambda_h \lambda_p + 8 \lambda_p \lambda_s+4 \lambda_p^2+6 \lambda_p y_t^2 + 4 N_f y_\psi^2 \lambda_p  \,,\\
(4 \pi)^2 \beta_{y_t} &=  - \tilde f_y y_t - \frac{17}{12} g^2 y_t - \frac{2}{3} g_D^2 y_t - \frac{5}{3} g_D g_\epsilon y_t - \frac{9}{4} g_2^2 y_t - 8 g_3^2 y_t-\frac{17}{12} g_\epsilon ^2 y_t+\frac{9}{2} y_t^3 \,, \\
(4 \pi)^2 \beta_{y_\psi} &= - \tilde f_y y_\psi  + 6 y_\psi^3  - 6 g_D^2 y_\psi \,.
\end{align}
Here, $N_f$ refers to the degrees of freedom counted in Weyl fermions of the vectorlike fermion $\psi$.

\subsection{\texorpdfstring{$U(1)_X$}{U(1)X} model}
The beta functions in the $U(1)_X$ model are given by
\begin{align}
(4 \pi)^2 \beta_{g_D} &= -\tilde f_g g_D+ (4 n_\psi^2 + \frac{1}{3}n_S^2) g_D^3+\frac{41}{6} g_D g_\epsilon ^2 \,,\\
(4 \pi)^2 \beta_{g_\epsilon } &=  -\tilde f_g g_\epsilon + (4 n_\psi^2 + \frac{1}{3}n_S^2)  g_D^2 g_\epsilon + \frac{41}{6} g_\epsilon ^3 + \frac{41}{3} g^2 g_\epsilon \,,\\
(4 \pi)^2 \beta_{\lambda_h} &=  \tilde f_\lambda \lambda_h+\frac{3}{8} g^4+\frac{3}{4} g^2 g_2^2+\frac{3}{4} g^2 g_\epsilon ^2-3 g^2 \lambda_h+\frac{9}{8} g_2^4 + \frac{3}{4} g_2^2 g_\epsilon ^2 -9 g_2^2 \lambda_h+\frac{3}{8} g_\epsilon ^4 - 3 g_\epsilon ^2 \lambda_h + 24 \lambda_h^2 - 6 y_t^4 +12 \lambda_h y_t \,,\\
(4 \pi)^2 \beta_{y_t} &=   -\tilde f_y y_t-\frac{17}{12} g^2 y_t  - \frac{9}{4} g_2^2 y_t - 8 g_3^2 y_t - \frac{17}{12} g_\epsilon ^2 y_t+\frac{9}{2} y_t^3 \,.
\end{align}
The model can be extended with an additional Higgs portal scalar.
The beta functions of $\lambda_p$ and $\lambda_S$ as well as their contributions to $\beta_{\lambda_h}$ are then the same as in the $B$-$L$ model.
The contributions to the gauge couplings are displayed via the quantum number $n_S$.
\end{widetext}

\footnotesize
\bibliographystyle{abbrv}



\end{document}